# Photon super-bunching from a generic tunnel junction

Christopher Leon[*,†,=], Anna Rosławska[†,=], Abhishek Grewal[†], Olle Gunnarsson[†], Klaus Kuhnke[†], Klaus Kern[†,‡]

[†]Max-Planck-Institut für Festkörperforschung, Heisenbergstraße 1, DE-70569 Stuttgart, Germany.

[‡]Institut de Physique, École Polytechnique Fédérale de Lausanne, CH-1015 Lausanne, Switzerland.

[*]Corresponding author: c.leon@fkf.mpg.de

[=]Equal contribution

**Generating correlated photon pairs[1] at the nanoscale is a prerequisite to creating highly integrated optoelectronic circuits that perform quantum computing tasks[2] based on heralded single-photons[3]. Here we demonstrate fulfilling this requirement with a generic tip-surface metal junction. When the junction is luminescing under DC bias, inelastic tunneling events of single electrons produce a photon stream in the visible spectrum whose super-bunching index is 17 when measured with a 53 picosecond instrumental resolution limit. These photon bunches contain true photon pairs of plasmonic origin, distinct from accidental photon coincidences. The effect is electrically rather than optically driven – completely absent are pulsed lasers, down-conversions, and four-wave mixing schemes[4]. This discovery has immediate and profound implications for quantum optics and cryptography[5], notwithstanding its fundamental importance to basic science and its ushering in of heralded photon experiments[6] on the nanometer scale[7].**

Tunnel junctions are important light sources in their own right that convert electric potential energy into photons, largely though one-electron–one-photon (1e⁻ → 1γ) inelastic tunneling events. These junctions facilitate many intricate fundamental processes such as: correlated two-electron tunneling[8,9]; overbias emission[10,11,12]; photon anti-bunching in single-photon emitting molecular systems[13,14]; and photon bunching from dynamical processes that modulate junction properties, such as molecular motion[15,16]. These emission processes arise from how stochastic fluctuations couple to the electromagnetic modes of an environment[17], which imprint characteristic deviations away from Poissonian statistics onto the temporal correlations in the emitted photon stream. Using scanning tunneling microscopy (STM) induced luminescence techniques[18] to examine the light from atomically flat metal junctions, we observe a non-Poissonian process that manifests as photon super-bunching, and evidences emission of correlated photon pairs from a tunnel junction formed between any two metals. The effect is reminiscent of two-mode squeezed photon pairs[19], but without externally applied AC voltages and the energy constraints imposed by millikelvin temperatures.

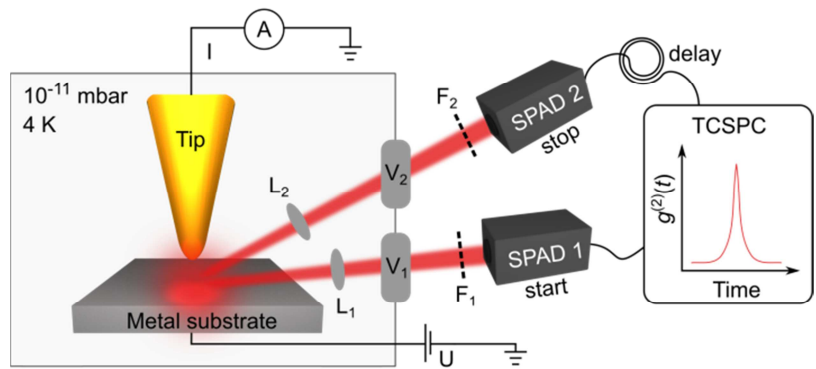

**Figure 1: Schematic of a scanning tunneling microscope combined with a Hanbury Brown and Twiss interferometer.** Light radiating from a junction formed between a Au tip and Ag(111) substrate travels along two optical paths (1, 2) through a series of (L)enses, (V)iewports, and optical (F)ilters to a pair of single-photon avalanche detectors (SPADs). The number of photon coincidence events as a function of time delay $t$ between the SPADs, $g^{(2)}(t)$, is measured with a time-correlated single-photon counter (TCSCP). Voltage bias (U) is applied to the substrate. The tunnel current (I) is measured with a picoammeter (A). A third optical path to an optical spectrometer is not shown.

This Letter reports on the measurements obtained with the experimental setup[20] shown in Figure 1. The surface topography and spectroscopic characterization of a clean Ag(111) single crystal obtained with STM are shown in Figure 2a and Figure 2b, respectively. Light radiating from the junction (orange curve) due to the tunnel current is recorded while sweeping the bias from 1 V to 10 V, holding the current constant with a feedback loop. The feedback causes the tip to retract from the surface in a step-like fashion (purple curve) due to field emission resonance states (FER, green curve) at metal surfaces with a band gap near the vacuum level[21]. The succession of these FER states introduces oscillatory variations in the electronic density of states. Note that the total light emission intensity (orange curve) drops significantly from its maximum near 3 V when the voltage approaches the first FER state. Our measurement reproduces the essential, known features of a metal-metal tunnel junction[22,23].

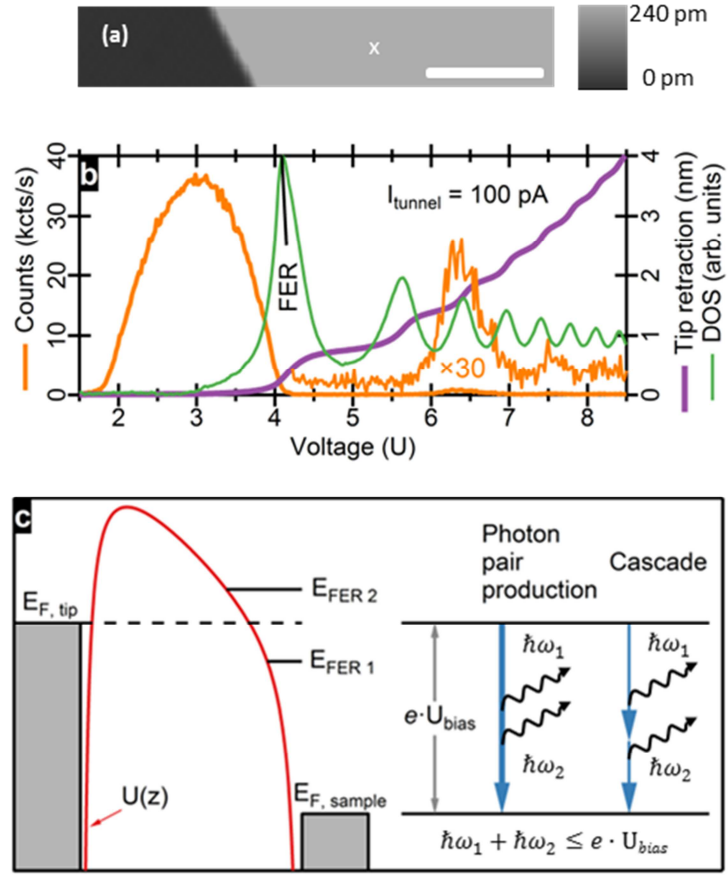

**Figure 2: Tunnel junction characterization with photon pair generation schematic. a**, Ag(111) surface topography with a monatomic step imaged at 3V, 100 pA. X marks the position of bunching measurements. Scale bar is 5 nm long. The gradation spans one 240 pm Ag terrace step height. **b**, Total light intensity (orange), tip retraction (purple), and density of states (green, DOS) during a linear voltage sweep at constant current. The position of the first field emission resonance (FER) maximum is indicated with a black line. **c**, An energy level diagram of an inelastic electron tunneling event leading to photon pair production. The junction is biased by $U_{bias}$ voltage. An electron at the tip Fermi level ($E_{F,\,tip}$) tunnels through a junction potential barrier $U(z)$, arriving on the sample side with an energy $E^* = e \cdot U_{bias}$ above the sample Fermi level ($E_{F,\,sample}$). $E^*$ can be aligned or misaligned with FERs nearby.

The temporal photon intensity correlation function $g^{(2)}(t)$ that evidences photon super-bunching is measured with a Hanbury Brown and Twiss interferometer[24] (Figure 1) by collating the distribution of times $t$ between one photon arriving at the start detector (SPAD 1) and another photon arriving at the stop detector (SPAD 2)[13]. Two photon counters are necessary to confirm simultaneously generated photons because the instrumental dead time is ~76 ps (see Methods). A correlation event registers when both detectors sense one photon each, typically with a nanosecond time delay between the sensing. While accidental coincidences may occur at any relative time delay (as they involve uncorrelated photons), true coincidences require two emitted photons arriving simultaneously, and can only manifest as a sharp feature in $g^{(2)}(t)$ at time-zero. These special pairs can be produced according to the schematic shown in Figure 2c. An inelastic tunneling process excites some plasmon modes that subsequently decay into photons detected in the far field. In addition to well-known single-photon emission, bunched emission may occur in a single step producing a photon pair, or two-step cascade[25] that produces one photon in each step.

Figure 3a shows the measured $g^{(2)}(t)$ for our tunnel junction light source operated at 4.63 V, 20 nA derived from time-correlated single-photon counting and plotted with coincidence events as a function of time between photon detection at the start and stop SPADs. While observing $g^{(2)}(0) > 1$ is already indicative of bunched photon emission, $g^{(2)}(0) = 17$ (Figure 3a) shows that the photons are unambiguously super-bunched[26]. Importantly, the instrumental response function must dominate the bunching feature because its width is even narrower than that attained with reference picosecond light pulses (blue curve in Figure 3b; Methods). Thus, the peak value of $g^{(2)}(0)$ is limited by the detectors' time resolution. Using $g^{(2)}(0)$ as a coincidence-to-accidental ratio, this metric is already comparable to photon pair sources based on cooled optical fibers, which can perform quantum key distribution with a 3% bit error rate[27].

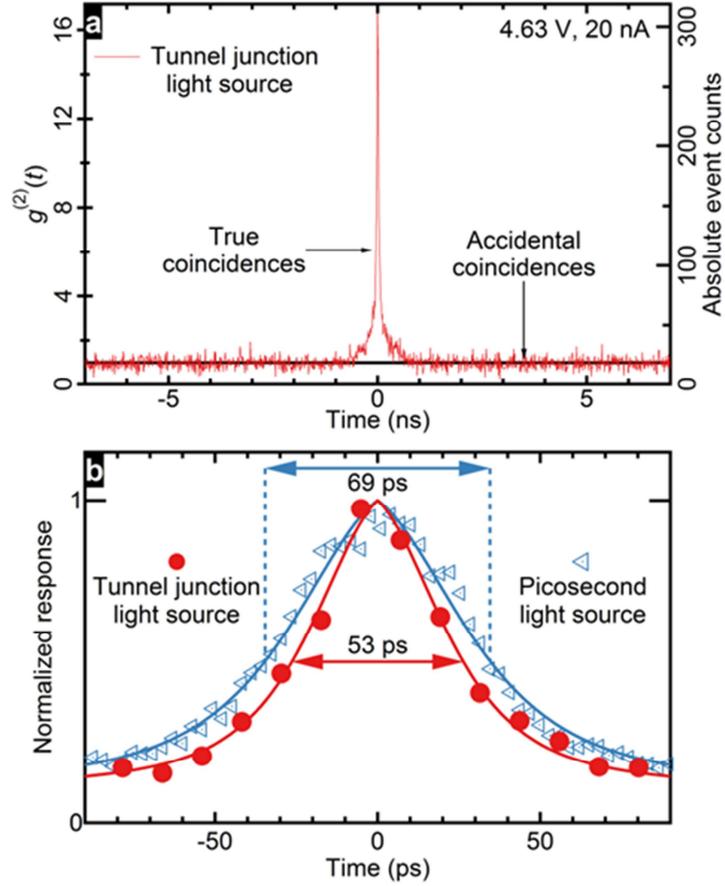

**Figure 3: Photon correlation measurements of a tunnel junction and picosecond light source. a**, Typical $g^{(2)}(t)$ measurement for the tunnel junction light source. The total number of true coincidence events (2620) is determined by integrating between ±7 ns after subtracting the level of accidental coincidence events (18.6) that corresponds to $\lim_{t \to \infty} g^{(2)}(t) = 1$ (black line) and is equal to the product of the two SPAD count rates. Total accumulation time of 29400 s. **b**, A comparison of $g^{(2)}(t)$ rescaled to have unity peak height for the tunnel junction source (red) and an autocorrelation of a commercial picosecond white light source with 6 ps fundamental pulsewidth (blue). The full widths at half the maxima are indicated. Solid lines are guides for the eye.

Next we characterize how the true and accidental coincidence events vary as a function of tunnel current. The raw data and resulting series is shown in Figure 4. For each new current, the bias is adjusted slightly to follow the first FER maximum because reproducing the measuring conditions at the FER is facile[23]; other well-defined setpoints are equally valid. Bunching is counterintuitively best observed when the total light intensity is made low [28], either by reducing the tunnel current (Figure 3a and Figure 2a(i)), or in tandem, leveraging the broad, low but non-zero minimum between 4 V and 6 V in the light intensity curve (Figure 2b, orange). Bunching does not seem to be predicated on populating the FER states because it is visible at 3.2 V which is below the first FER maximum.

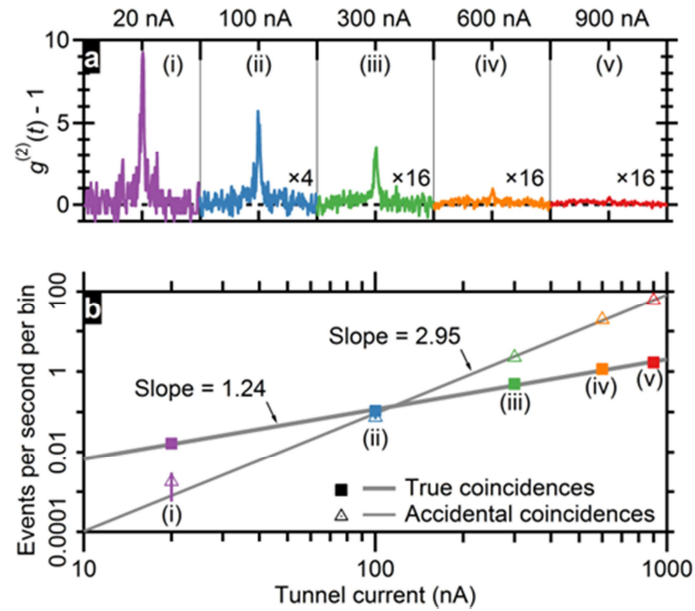

**Figure 4. Photon correlation measurements as a function of current. a**, The measured function $g^{(2)}(t) - 1$, labeled with consecutive Roman numerals. The unity shift aligns the normalized accidental coincidence level to zero across all measurements for ease of comparison. The horizontal axis in each window spans ±1 ns time delay *t* with the current and applied voltage shown. Voltages of 4.47, 4.60, 4.65, 4.70, and 4.74 V applied for the data in (i-v) respectively. **b**, log-log plot of the absolute number of true and accidental coincidence events versus current. Both have a power-law dependence on the current equal to the slope of the fitted lines. The ratio of each data point pair gives the respective $g^{(2)}(0) - 1$ peak values of (i) 9.2, (ii) 1.4, (iii) 0.22, (iv) 0.060, and (v) 0.028 for the traces in **a**. Total accumulation time of 1200 s (i-iii) and 600 s (iv-v).

Figure 4a shows the measured quantity $g^{(2)}(t) - 1$. From right to left, the bunching peak value increases with decreasing current. Focusing on Figure 4b, the power law exponent for true coincidences (1.24) is less than half of the value for accidental coincidences (2.95), implying that bunching is due to single electron tunneling events, and for this reason it will dominate at low current. Let $I$ be the current, $e$ the electric charge, $k_1$ the quantum efficiency for emitting a single photon, and η the effective probability that this photon is detected. Without loss of generality, η for both detectors is made equal. The number of single-photon events per second is $N_1 = Ik_1\eta/e$, hence, the number of accidental coincidences per second is $\widehat{N}_2 = N_1^2\tau = (Ik_1\eta/e)^2\tau$, where τ is the binning time interval of the correlation measurement. In contrast, the number of true coincidence events per second is $N_2 = Ik_2\eta^2/e$, where $k_2$ is the quantum efficiency for emitting a photon pair. We assume that $k_2 \ll k_1 \ll 1$ so mistaking photon pairs for single photons is negligible. The ratio of correlated pairs to accidental coincidences is then $g^2(0) - 1 \approx \frac{N_2}{\widehat{N}_2} = \left(\frac{e}{I\tau}\right)\left(\frac{k_2}{k_1^2}\right)$. If the photon pair originates from a cascade of two independent emission events with the same quantum efficiency, $k_2 = k_1^2$. Actually, we find that $k_2 = 215k_1^2$ (Figure 3a) can be much larger than $k_1^2$. If the tunneling electron loses energy in an emission process, it tends to fall into a state which, in the classically forbidden region, has an exponentially smaller amplitude than the initial state. This should reduce the probability for a second photon emission. Calculations suggest a reduction by a few orders of magnitude. The observed $k_2$ is then several orders of magnitude larger than expected for a cascade of two independent emission events. Furthermore, we notice different current dependencies of $k_1^2$ and $k_2$ (Figure 4b); $k_1 \sim I^{0.475}$ and $k_2 \sim I^{0.24}$ rather than $k_2 \sim k_1^2 \sim I^{0.95}$, as might have been expected for a cascade process. These two observations raise fundamental questions about the nature of two-photon processes. They are also two strong arguments against a cascade emission mechanism and both evidence a coherent simultaneous pair emission process being operant instead.

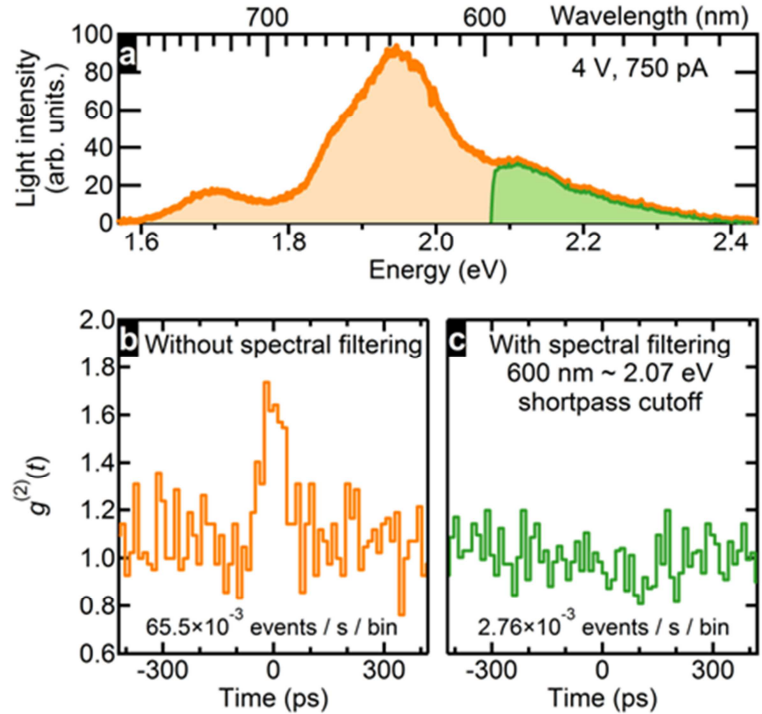

**Figure 5: Photon correlation measurements at fixed tunnel conditions with varied spectral filtering. a**, Measured optical spectrum (orange) and its shortpass 600 nm cutoff spectrum (green). **b**, Bunching is observed in the unfiltered light. **c**, Bunching is *not* observed when the low energy photons are blocked for both detectors, and the total energy of a photon pair is required to exceed the electron energy of 4 eV. The accidental correlation event rate is shown for **b** and **c**. Total accumulation time is 600 s and 39000 s, respectively.

As a consistency check, $g^{(2)}(t)$ is measured with and without intervening optical shortpass filters ($F_1$ and $F_2$ in Figure 1) with cutoffs exceeding half the applied tunnel voltage. This tests whether positive correlations occur for photon pairs whose total energy exceeds the energy of one tunneling electron. Figure 5 shows the light spectrum obtained with $e \cdot U$ = 4 eV tunneling electrons which exhibits bunching (orange; parts a, b), while the filtered spectrum with a 2.07 eV cutoff does not (green; parts a, c). Hence, in a bunch (or pair) of photons, not more than one photon carries more than half of the electron energy.

This result reaffirms that the bunched photons do not originate from sequential $1e^- \rightarrow 1\gamma$ processes. Even bunching from coordinated electron tunneling is unlikely because this process is quadratic in the current[29]. We speculate that photon pair creation is promoted by the small volume and spatial asymmetry of the tunnel junction, and the absence of optical resonances in the junction (spectrum in Figure 5a) whose energies overlap with the distribution of tunneled electron energies.

Confirming the existence of a $1e^- \rightarrow 2\gamma$ electroluminescent process in the Au-Ag(111) junction is the key to rationalizing our bunching observations. In fact, such a process evidences photon pairs being emitted much closer to each other than the ~50 ps temporal resolution of our experiments. This result validates that simple tunnel junctions do behave as special light emitters beyond their usual domain. We have also observed bunching from other junctions, notably Au-Cu(111). Using quantum efficiency and energy conservation arguments, we demonstrated that photon pairs can originate from an intrinsic elementary process with fast temporal characteristics without invoking bulk optical media with strong $2^{nd}$ or $3^{rd}$ order nonlinearities.

Our results show that an optoelectronic component useful for quantum computing can be miniaturized significantly and controlled at the atomic scale (Figure 1a). Nanoscale photon entanglement and heralding measurements which require double- or higher-photon coincidences may now be possible; the coincidence rate in Figure 3a is already sufficient to accomplish the former[30]. Using broader tip apexes[31], the photon energy may shift to where optical fibers have minimal transmission loss (C-band, 1530-1565 nm) and synergize with contemporary developments in optical communications[32]. We anticipate these findings will further motivate using tunnel junctions as novel photonic devices in nanoscience.

# Methods

**Peripherals.** The scanning tunneling microscopy induced luminescence setup is equipped with a pair of spectrally integrating, single-photon avalanche photodiodes (SPAD, Micro Photon Devices PD-100-CTE, supplier-measured 29 ps and 33 ps FWHM time resolution, 74.9 ns and 76.8 ns dead time) and an optical spectrograph (Acton Research Spectra Pro 900i; 150 lines/mm blazed grating with Peltier-cooled, intensified charge-coupled device). In all optical spectra shown, no correction is made for variations in the wavelength dependence of the detection efficiency. Photon correlation measurements are obtained with hardware (Becker & Hickl, SPC-130) configured in start-stop mode with a 12.2 ps time bin width throughout this study. A white light source (Fianium WhiteLase, WL-SC-400-40, 80 MHz repetition rate, 6 ps fundamental pulsewidth) filtered by a variable bandpass monochromator (Fianium LLTF Contrast VIS, < 2.5 nm spectral bandwidth) generates the 690 nm pulses used in the reference measurement shown in Figure 3b.

**Tunnel junction.** A Ag(111) single crystal, oriented to 0.1°, is cleaned via repeated cycles of $Ar^+$ sputtering between 300–400 K followed by annealing to 900 K in an ultra-high vacuum preparation chamber. It is then transferred *in situ* to the scanning tunneling microscope and checked for surface cleanliness (Figure 2a). Tips are prepared by electrochemically etching ⌀0.25 mm, 99.995% pure Au wire, followed by repeated indenting and voltage pulsing on Ag(111). Differential conductance (dI/dU) measurements, in which the signal is proportional to the density of electronic states, are made by modulating the bias voltage (211 Hz, 20 mV peak-to-peak) and recording the lock-in signal in the current. All voltages are applied to the Ag substrate with the tip held at 0 V. All measurements with the junction are reproducibly tunable via STM parameters such as voltage and current excluding any tip modification.

## Author Contributions

Experimental work done by CL, AR, AG. Comparison with the cascade emission model done by OG. CL analyzed the data and wrote the manuscript with input from all authors. KKu and KKe conceived of and designed the research program.

## Competing Interests

The authors declare no competing financial interests.

## References

[1] Orieux, A., Versteegh, M. A. M., Jöns, K. D. & Ducci, S. Semiconductor devices for entangled photon pair generation: a review. *Rep. Prog. Phys.* **80**, 076001 (2017).

[2] Knill, E., Laflamme, R. & Milburn, G. A scheme for efficient quantum computation with linear optics. *Nature* **409**, 46–52 (2001).

[3] Collins, M. J. *et al.* Integrated spatial multiplexing of heralded single-photon sources. *Nat. Commun.* **4**, 2582 (2013).

[4] Eisaman, M. D., Fan, J., Migdall, A. & Polyakov, S. V. Invited review article: single-photon sources and detectors. *Rev. Sci. Instrum.* **82**, 071101 (2011).

[5] Jennewein, T., Simon, C., Weihs, G., Weinfurter & H., Zeilinger, A. Quantum cryptography with entangled photons. *Phys. Rev. Lett.* **84**, 4729-4732 (2000).

[6] Wagenknecht, C. *et al*. Experimental demonstration of a heralded entanglement source. *Nat. Photonics* **4**, 549-552 (2010).

[7] Gentry, C. M. *et al*. Quantum-correlated photon pairs generated in a commercial 45 nm complementary metal-oxide semiconductor microelectronic chip. *Optica* **2**, 1065-1071 (2015).

[8] Kaasbjerg, K. & Nitzan, A. Theory of light emission from quantum noise in plasmonic contacts: above-threshold emission from higher-order electron-plasmon scattering. *Phys. Rev. Lett.* **114**, 126803 (2015).

[9] Xu, F., Holmqvist, C., Rastelli, G. & Belzig, W. Dynamical Coulomb blockade theory of plasmon-mediated light emission from a tunnel junction. *Phys. Rev. B* **94**, 245111 (2016).

[10] Dong, Z. C. *et al*. Generation of molecular hot electroluminescence by resonant nanocavity plasmons. *Nat. Photonics* **4**, 50-54 (2010).

[11] Peters, P.-J. *et al*. Quantum coherent multielectron processes in an atomic scale contact. *Phys. Rev. Lett.* **119**, 066803 (2017).

[12] Schull, G., Néel, N., Johansson, P. & Berndt, R. Electron-plasmon and electron-electron interactions at a single atom contact. *Phys. Rev. Lett.* **102**, 057401 (2009).

[13] Merino, P., Große, C., Rosławska, A., Kuhnke, K. & Kern, K. Exciton dynamics of $C_{60}$-based single-photon emitters explored by Hanbury Brown-Twiss scanning tunneling microscopy. *Nat. Commun.* **6**, 8461 (2015).

[14] Zhang, L. *et al*. Electrically driven single-photon emission from an isolated single molecule. *Nat. Commun.* **8**, 580 (2017).

[15] Silly, F. & Charra, F. Time-autocorrelation in scanning-tunneling-microscope-induced photon emission from metallic surface. *Appl. Phys. Lett.* **77**, 3648 (2000).

[16] Perronet, K., Schull, G., Raimond, P. & Charra, F. Single-molecule fluctuations in a tunnel junction: a study by scanning-tunnelling-microscopy-induced luminescence. *Europhys. Lett.* **74**, 313-319 (2006).

[17] Jin, J., Marthaler, M. & Schön, G. Electroluminescence and multiphoton effects in a resonator driven by a tunnel junction. *Phys. Rev. B* **91**, 085421 (2015).

[18] Kuhnke, K., Große, C., Merino, P. & Kern, K. Atomic-scale imaging and spectroscopy of electroluminescence at molecular interfaces. *Chem. Rev.* **117**, 5174-5222 (2017).

[19] Forgues, J-C., Lupien, C., Reulet, B. Emission of microwave photon pairs by a tunnel junction. *Phys. Rev. Lett.* **113**, 043602 (2014).

[20] Kuhnke, K. *et al*. Versatile optical access to the tunnel gap in a low temperature scanning tunneling microscope. *Rev. Sci. Instrum.* **81**, 113102 (2010).

[21] Chulkov, E. V. *et al*. Electronic Excitations in Metals and at Metal Surfaces. *Chem. Rev.* **106**, 4160-4206 (2006).

[22] Berndt, R. & Gimzewski, J. K. Isochromat spectroscopy of photons emitted from metal surface in an STM. *Ann. Physik* **2**, 133-140 (1993).

[23] Martínez-Blanco, J. & Fölsch, S. Light emission from Ag(111) driven by inelastic tunnelling in the field emission regime. *J. Phys.: Condens. Matter* **27**, 255008 (2015).

[24] Hanbury Brown, R. & Twiss, R. Q. Interferometry of the Intensity Fluctuations in Light II. An experimental test of the theory for partially coherent light. *P. Roy. Soc. Lond. A Mat.* **243**, 291-319 (1958).

[25] Zhang, X., Chang, X. & Ren, Z. A simple and general strategy for generating frequency-anticorrelated photon pairs. *Sci. Rep.* **6**, 24509 (2016).

[26] McNeil, K. J. & Walls, D. F. Possibility of observing enhanced photon bunching from two photon emission. *Phys. Lett. A* **51**, 233-234 (1975).

[27] Takesue, H. & Inoue, K. 1.5-μm band quantum-correlated photon pair generation in dispersion-shifted fiber: suppression of noise photons by cooling fiber. *Opt. Express* **13**, 7832-7839 (2005).

[28] González-Tudela, A., del Valle, E. & Laussy, F. P. Optimization of photon correlations by frequency filtering. *Phys. Rev. A* **91**, 043807 (2015).

[29] Xu, F., Holmqvist, C. & Belzig, W. Over-bias light emission due to higher order quantum noise of a tunnel junction. *Phys. Rev. Lett.* **113**, 066801 (2014).

[30] Franson, J. D. Two-photon interferometry over large distances. *Phys. Rev. A* **44**, 4552-4555 (1991).

[31] Boyle, M. G., Mitra, J. & Dawson, P. Infrared emission from tunneling electrons: The end of the rainbow in scanning tunneling microscopy. *Appl. Phys. Lett.* **94**, 233118 (2009).

[32] Agrell, E. *et al*. Roadmap of optical communications. *J. Opt.* **18**, 063002 (2016).